\documentclass[conference]{IEEEtran}
\IEEEoverridecommandlockouts
\usepackage{cite}
\usepackage{amsmath,amssymb,amsfonts}
\usepackage{graphicx}
\usepackage{textcomp}
\usepackage{xcolor}
\usepackage{multirow}
\usepackage{graphicx}

\usepackage{pifont}
\newcommand{\xmark}{\ding{55}}%
\usepackage{subcaption}
\usepackage[dvipsnames]{xcolor}

\usepackage{adjustbox}
\usepackage{amsmath,amssymb,amsfonts, amsthm}
\usepackage[linesnumbered,ruled,vlined]{algorithm2e}
\usepackage[noend]{algpseudocode}
\usepackage{tikz}
\usepackage{dblfloatfix}
\usepackage{fancyhdr}

\def\BibTeX{{\rm B\kern-.05em{\sc i\kern-.025em b}\kern-.08em
    T\kern-.1667em\lower.7ex\hbox{E}\kern-.125emX}}

\usepackage[most]{tcolorbox}

\newtcbtheorem{Summary}{\bfseries Summary}{enhanced,drop shadow={black!50!white},
  coltitle=black,
  top=0.3in,
  attach boxed title to top left=
  {xshift=1.5em,yshift=-\tcboxedtitleheight/2},
  boxed title style={size=small,colback=pink}
}{summary}

\newtcolorbox[auto counter]{summary}[1][]{title={\bfseries Summary~\thetcbcounter},enhanced,drop shadow={black!50!white},
  coltitle=black,
  top=0.3in,
  attach boxed title to top left=
  {xshift=1.5em,yshift=-\tcboxedtitleheight/2},
  boxed title style={size=small,colback=pink},#1}

\begin{document}

\title{An Automated Framework for Generating Stealthy Cell-Embedded Hardware Trojans\\ 
}

\author{
\IEEEauthorblockN{Raghul Saravanan\IEEEauthorrefmark{1}, Sudipta Paria\IEEEauthorrefmark{2}, Sai Manoj P D\IEEEauthorrefmark{1} and Swarup Bhunia\IEEEauthorrefmark{2}}
\IEEEauthorblockA{\IEEEauthorrefmark{1}Department of Electrical and Computer Engineering, George Mason University, Fairfax, VA}
\IEEEauthorblockA{\IEEEauthorrefmark{2}Department of Electrical and Computer Engineering, University of Florida, Gainesville, FL\\
rsaravan@gmu.edu, sudiptaparia@ufl.edu, spudukot@gmu.edu, swarup@ece.ufl.edu}
}

\maketitle

\begin{abstract}
Hardware Trojans (HTs) pose significant threats across the Integrated Circuit (IC) design lifecycle because they can be inserted by untrusted entities at different stages under the zero-trust model. When triggered under rare conditions, HTs can compromise the functionality, reliability, or security of the fabricated chip. HT assessment is typically performed by modeling realistic Trojan insertion scenarios 
in RTL implementation or gate-level netlists. While this model is useful for evaluating detection methods, it does not capture attacks where malicious behavior is hidden inside standard-cell implementations from a compromised library supplied by an untrusted vendor. This paper presents a novel framework for automatically generating cell-embedded hardware Trojans using compromised standard-cell implementations. Our proposed framework analyzes a mapped design, identifies candidate cell instances with rare input conditions, and applies payload templates that corrupt the selected cell output only when the trigger condition is satisfied. Experiments on open-source combinational and sequential benchmark designs show that our proposed framework can generate valid and stealthy Trojan instances across different cell types and design sizes. The results highlight a critical gap in current Trojan detection assumptions and show the need for cell-aware validation of standard-cell implementations in zero-trust IC design flows.

\end{abstract}


\section{Introduction}

\begin{figure}[!ht]
    \centering
    \includegraphics[width=0.95\columnwidth]{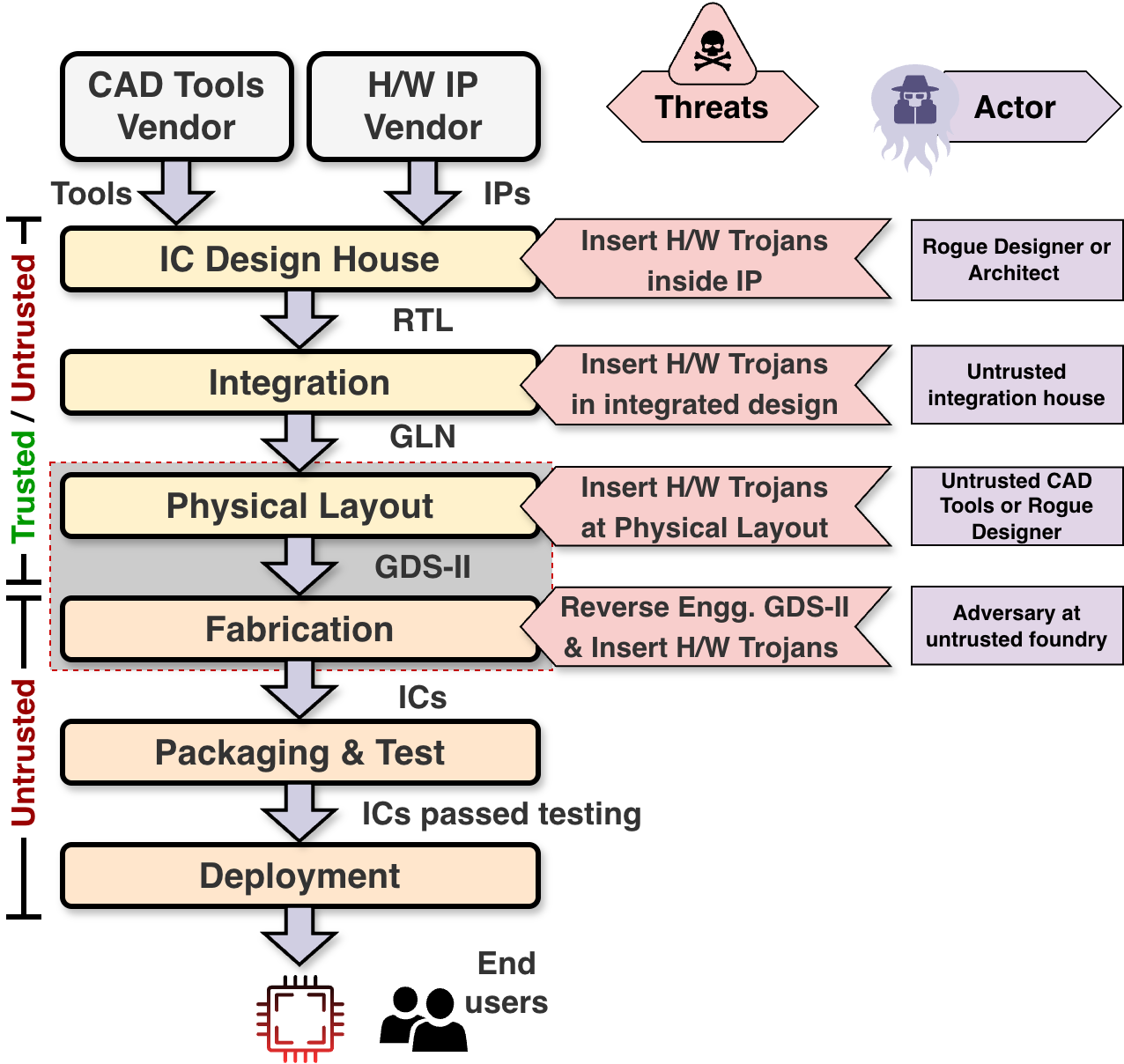}
    \caption{Hardware Trojan insertion at different stages of IC life cycle under the zero-trust model.}
    \label{fig:threat_model}
\end{figure}

Hardware Trojan (HT) attacks pose a significant and growing threat to the security and integrity of integrated circuits (ICs). Trojans can be inserted by untrusted parties at various stages of the IC design flow under the zero-trust model, as depicted in Fig.~\ref{fig:threat_model}. Traditionally, HTs are modeled as explicit modifications to RTL or gate-level netlists by inserting additional trigger and payload logic \cite{book}. However, an adversary controlling the standard-cell library or untrusted foundry can instead compromise the implementation of library cells themselves, allowing malicious behavior to be introduced without altering the synthesized netlist \cite{synfuzz}. In such a scenario, the IP designer provides a trusted RTL design, while the adversary tampers with the standard-cell library before synthesis. The modified library is subsequently characterized and compiled into the timing library used by commercial synthesis tools. Since the compromised cells preserve their intended functionality during normal operation, the synthesis tool treats them as legitimate implementations and instantiates them during technology mapping without detecting the embedded malicious logic. As a result, Trojans become embedded within the final hardware implementation while the synthesized netlist remains structurally identical to that produced using a trusted library. Benchmarking efforts \cite{shakya2017benchmarking,trusthub,cadforassurance} addressed the need for standardized evaluation by providing collections of designs containing predefined Trojans. Automated frameworks such as \cite{trit} further generalize this process by dynamically inserting diverse Trojan structures into gate-level netlists using rare-net analysis. Nevertheless, these approaches remain restricted to inserting explicit trigger and payload circuitry at the RTL or gate-level abstractions and therefore cannot model attacks originating from compromised standard-cell libraries. Consequently, they fail to represent an increasingly realistic threat in which malicious behavior is hidden inside legitimate library cells supplied by an untrusted library vendor or foundry.

\begin{figure*}[!ht]
\centering
\subfloat[Original Circuit]{\includegraphics[width=0.66\columnwidth]{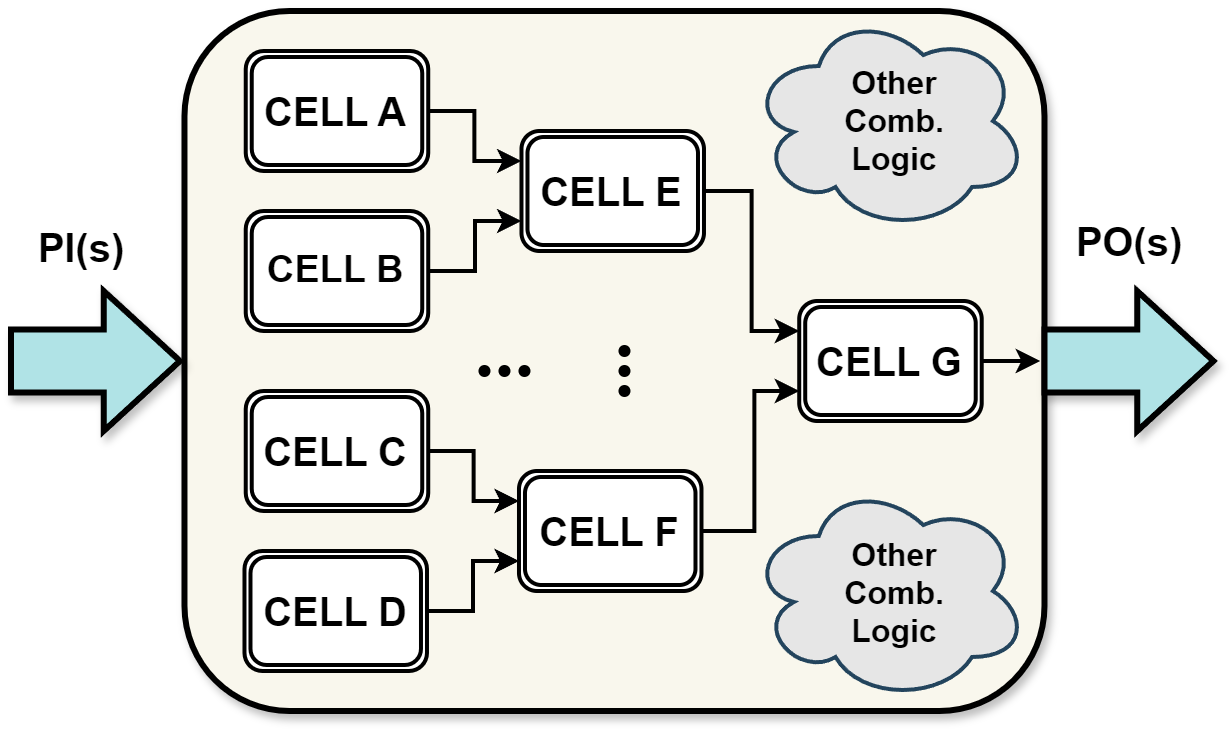}}
\hfill
\subfloat[Traditional HT Insertion]{\includegraphics[width=0.661\columnwidth]{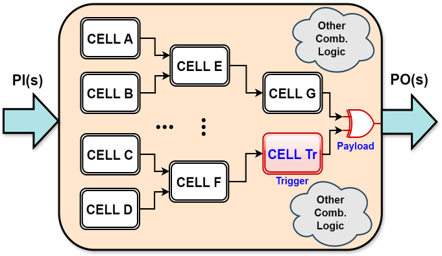}}
\hfill
\subfloat[Proposed HT Insertion]{\includegraphics[width=0.675\columnwidth]{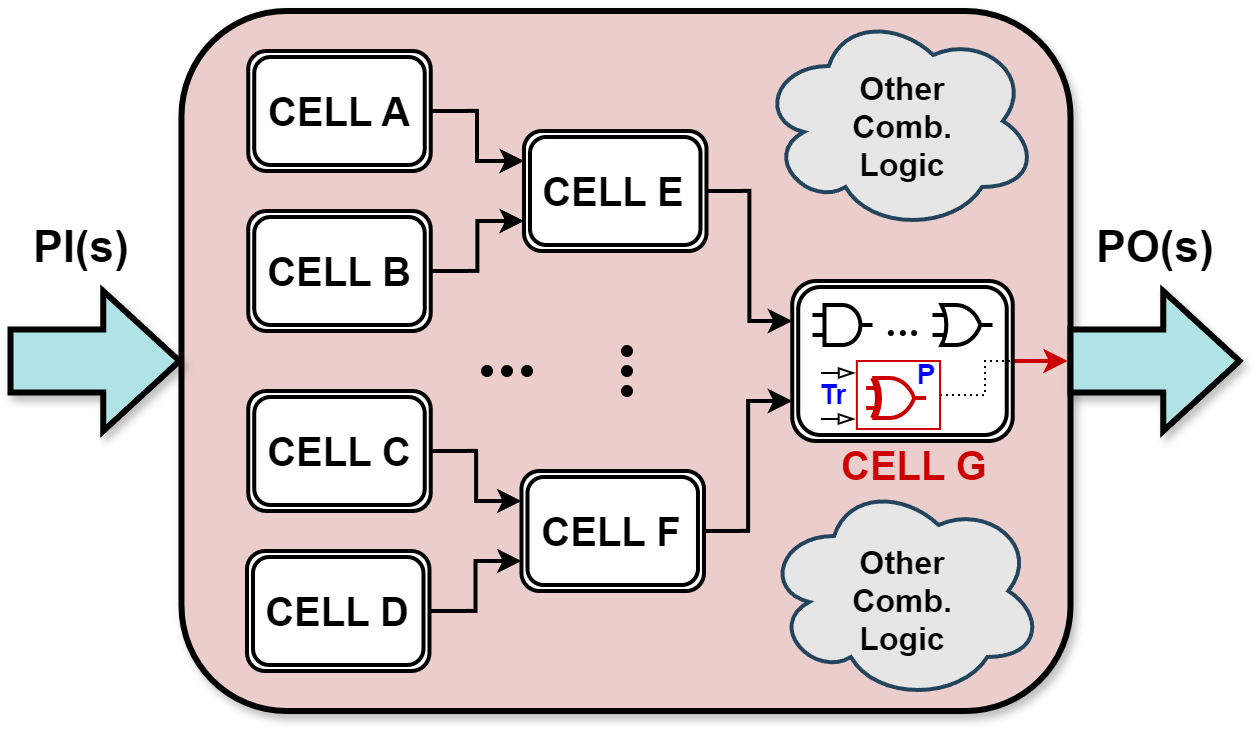}}
\caption{Comparison between conventional gate-level Trojan insertion and the proposed cell-embedded Trojan insertion model.}%
\label{fig:HT_insertion}%
\end{figure*}

Given the vast number of potential Trojans in even moderately complex ICs with diverse trigger conditions and payloads, hardware Trojan detection becomes a significant challenge. Consequently, HT detection has evolved toward scalable techniques including statistical test generation \cite{mero,latent,mero_2}, Boolean functional analysis \cite{func_analysis}, side-channel analysis \cite{side_channel_3,side_channel_2}, machine learning-based methods \cite{gaikwad2023hardware,ml_3}, and LLM-guided test generation \cite{ip_cots,covert} and detection \cite{Paria2026}. However, these techniques assume that the standard-cell library is trusted and operate on the synthesized gate-level netlist or physical implementation. Since a compromised library preserves the netlist structure while embedding malicious functionality within the internal implementation of selected cells, existing structural, statistical, and learning-based detection techniques cannot directly observe the inserted Trojan. This creates a critical blind spot in current hardware Trojan assessment methodologies and motivates the need for benchmark generation frameworks that explicitly model library-level attacks.

In this paper, we propose an automated framework for generating hardware Trojans by embedding configurable sequential trigger logic directly within standard-cell implementations. Following the zero-trust threat model in the traditional IC design flow, we assume an untrusted library provider or foundry capable of modifying the internal implementation of library cells before synthesis while preserving their normal functionality and characterization. Our proposed framework identifies candidate cells containing rare internal logic, embeds configurable sequence-detector-based Trojans inside their implementations, and produces Trojan-infected libraries that can be used by commercial synthesis tools without modifying the RTL or synthesized netlist. Fig.~\ref{fig:HT_insertion} compares conventional gate-level Trojan insertion with our proposed library-cell-based approach. Experimental results on a variety of benchmark designs demonstrate that the proposed framework generates valid and highly stealthy Trojans that evade representative state-of-the-art HT detection techniques, highlighting the importance of evaluating this emerging threat model.
In particular, this paper makes the following contributions:
\begin{itemize}
    \item We propose an automated framework for generating cell-embedded HTs by modifying selected standard-cell implementations while keeping the synthesized gate-level netlist structurally unchanged.
    \item We present a compromised-standard-cell threat model where an untrusted library provider or foundry modifies selected cell implementations for Trojan insertion.
    \item We evaluate the generated HTs on open-source designs and demonstrate low area/power overhead while evading common HT detection methods.
\end{itemize}

The remainder of this paper is organized as follows: Section II summarizes the relevant background. Section III describes the assumed threat model. Section IV outlines the methodology, and Section V contains the experimental results. We conclude the paper in Section VI. 

\section{Background}

In this section, we provide the necessary background on hardware Trojans, existing Trojan assessment techniques, and their limitations in modeling Trojans where malicious behavior is hidden inside compromised standard-cell implementations.

\subsection{Hardware Trojans}

\begin{figure}[!htbp]
\centering
\subfloat[]{\includegraphics[width=0.48\columnwidth]{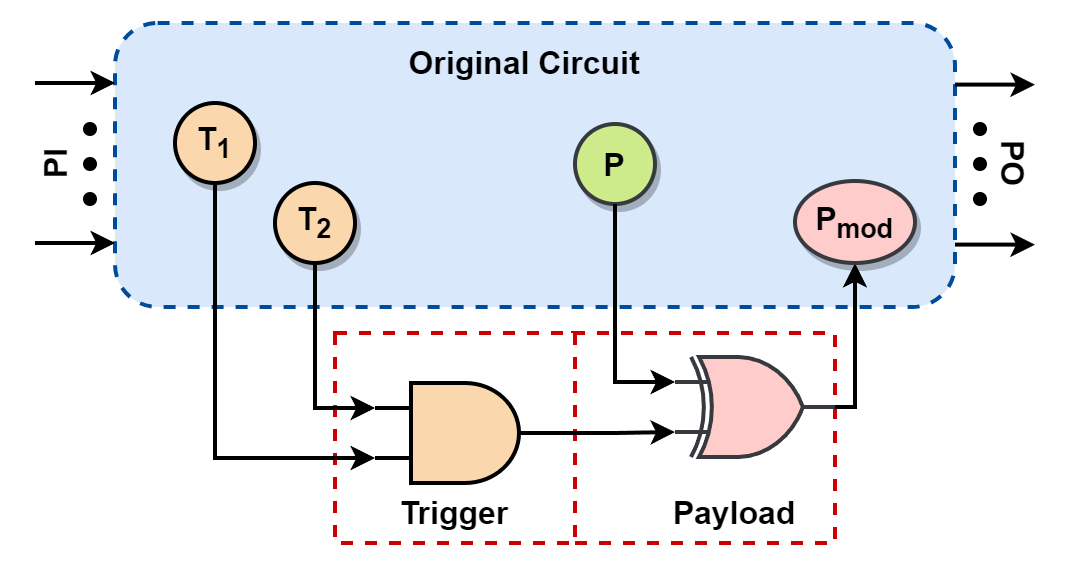}}
\hfill
\subfloat[]{\includegraphics[width=0.5\columnwidth]{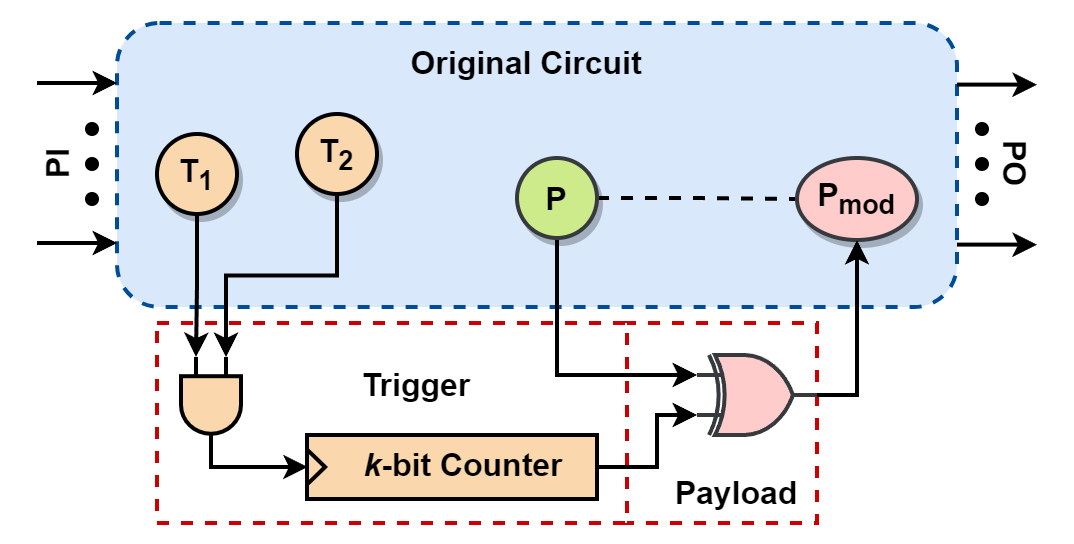}}
\caption{Generic Hardware Trojans: (a) a combinational Trojan that inverts the value of payload $P$ when $T1=T2=1$, and (b) a sequential Trojan that uses an asynchronous counter triggered by $T1$ and $T2$ that changes $P$ to $P_{mod}$ \cite{latent}.}%
\label{fig:Tr_comb_and_seq}%
\end{figure}

Hardware Trojans are intentional malicious modifications that potentially compromise the functionality, reliability, or security of an IC design \cite{tehranipoor2010survey,book}. A typical hardware Trojan consists of two main components: a trigger (activation) and a payload (effects). The trigger defines the condition under which the Trojan becomes active, while the payload defines the malicious effect after activation. The trigger can be always active, but most stealthy Trojans use rare internal signal values, rare input patterns, or long sequential events so that they remain inactive during normal validation and manufacturing test. This makes detection difficult because conventional simulation, random testing, and production test patterns may never activate the malicious behavior. Based on the trigger structure, Trojans are commonly grouped into \textit{combinational} and \textit{sequential} Trojans, as illustrated in Fig. \ref{fig:Tr_comb_and_seq}. A combinational Trojan is activated when a specific combination of internal signals occurs at the same time. A sequential Trojan requires a sequence of events across multiple clock cycles, often using counters, sequence detectors, finite state machines, or storage elements to track the activation condition.

\subsection{Existing HT Assessment Techniques}

Most existing HT benchmarks \cite{trusthub,cadforassurance} and insertion frameworks \cite{trit} model Trojans as explicit trigger and payload logic added to RTL or gate-level netlists. The inserted Trojans usually appear as additional logic in the design, even if the trigger condition is rare and difficult to activate. Consequently, several HT detection methods have been proposed to identify malicious modifications by activating rare-triggered Trojans.
Traditional random and constrained-random tests often fail to expose complex or rare-triggered Trojans, while directed tests offer targeted detection but struggle to scale with large designs. Side-channel analysis \cite{side_channel_3,side_channel_2} monitors physical parameters to detect anomalies, but it can be ineffective against subtle Trojans masked by process variations. Machine learning-based methods \cite{ml_1,ml_3,gaikwad2023hardware} leverage circuit features for Trojan classification but are typically performed at pre-silicon stages and trained on static benchmarks, limiting adaptability to real-world Trojan variants and the inability to detect Trojans that leverage compromised cells from untrusted foundries. Statistical test generation techniques \cite{mero,mero_2} are found to be promising alternatives to directed tests, which use the N-detect principle to activate rare nodes multiple times, aiming to uncover unknown triggers. The ATPG-based statistical test generation approach\cite{latent,tarmac_2} employs efficient test vector generation for HT detection, achieving high trigger and Trojan coverage for diverse designs. However, these techniques do not account for scenarios in which standard-cell libraries are compromised, such as when library cells obtained from an untrusted foundry contain embedded Trojans.

\subsection{Motivation}

Existing HT generation and detection techniques are found to be successful in HT assessment by inserting explicit Trojans (trigger+payload) into RTL or gate-level implementations. This does not fully capture attacks in which the malicious behavior is hidden inside the implementation of a standard cell belonging to a compromised library. In this scenario, the synthesized netlist may contain only legitimate cell instances, while corrupted cells behave incorrectly under rare input conditions. This creates a detection gap: test-generation and ML-based methods may not see additional suspicious logic, and side-channel methods may observe only small deviations. This motivates our current work which systematically generates Trojans hidden inside compromised cells and evaluates their triggerability, observability, and resistance against representative detection methods.

\section{Threat Model}

In this work, we assume a zero-trust IC design flow in which the standard-cell library provider or foundry is untrusted. The IP designer provides a trusted RTL or gate-level design, while the adversary compromises the standard-cell library before synthesis by modifying the implementation of selected library cells. Since the compromised cells preserve their intended functionality under normal operating conditions, the synthesis tool treats them as legitimate cells and instantiates them during technology mapping without detecting the embedded Trojan.

\noindent$\circ$ \textbf{Assets:} RTL and synthesized gate-level hardware IPs, together with the sensitive information processed or stored within the design.

\noindent$\circ$ \textbf{Adversary:} A malicious standard-cell library provider or untrusted foundry capable of modifying the implementation of standard cells.

\noindent$\circ$ \textbf{Adversarial Access:} Privileged access to physical layout,
EDA Tools, Cell Libraries GDS-II, netlist, and other related files

\noindent$\circ$ \textbf{Adversarial Objectives:} Embed stealthy hardware Trojans within selected standard-cell implementations while preserving their normal functionality. The Trojans remain dormant during normal operation and are activated only after observing rare internal trigger sequences, leading to denial-of-service or other malicious behaviors.

\noindent$\circ$ \textbf{Trust Model:} The IP owner/designer, RTL design, and synthesis flow are assumed to be trustworthy. The standard-cell library supplied to the synthesis tool is assumed to be potentially compromised.

\begin{figure}[!htbp]
\centering
\subfloat[Original Logic]{\includegraphics[width=0.43\columnwidth]{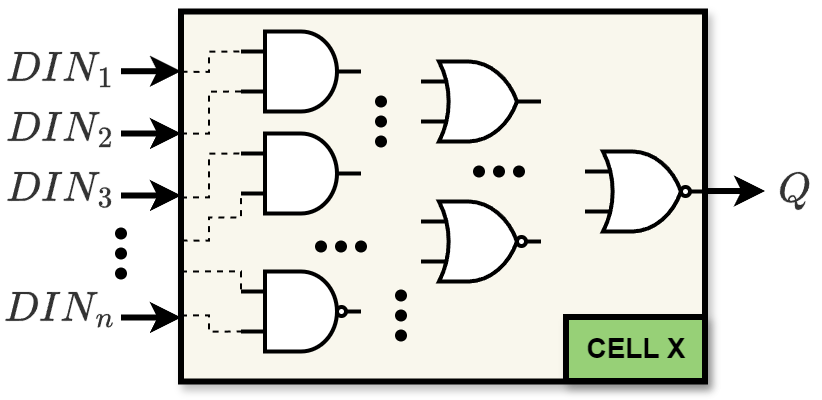}}
\hfill
\subfloat[Tampered Logic]{\includegraphics[width=0.56\columnwidth]{figs/tampered_cell.png}}
\caption{Example Standard Cell implementation with tampering.}%
\label{fig:cell}%
\end{figure}

\section{Proposed Framework: Trojan Generation and Validation}

\begin{figure*}[!htbp]
\centering
\includegraphics[width=\textwidth]{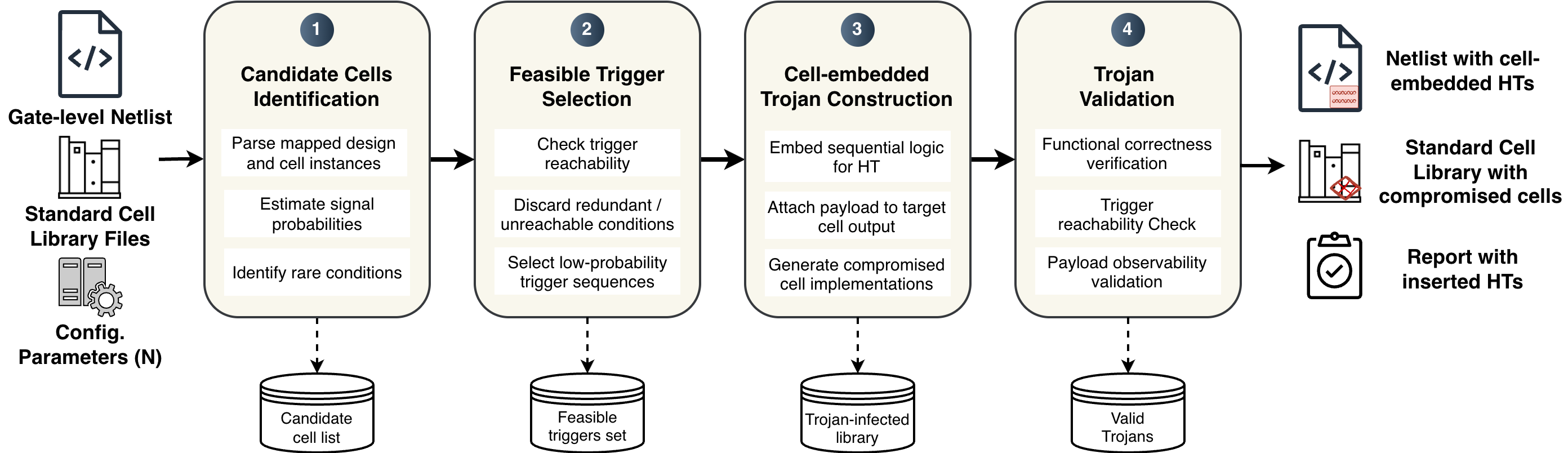}
\caption{Overview of our proposed HT insertion framework.}%
\label{fig:overview}%
\end{figure*}

The proposed framework generates cell-embedded hardware Trojans through a systematic four-stage flow, as shown in Fig. \ref{fig:overview}. Each of the four major stages of the proposed framework is described as follows.

\subsection{Identification of Candidate Cells}

The first stage identifies candidate standard-cell instances for Trojan insertion. Unlike conventional approaches that analyze rare nets at the gate-level netlist, our proposed framework examines the internal Boolean implementation of each mapped standard cell to identify rare input combinations that naturally occur with low probability during functional operation. Signal probabilities are propagated from the mapped design to estimate the activation likelihood of each cell input condition. Cell instances containing sufficiently rare internal logic become candidate locations for embedding stealthy hardware Trojans. This enables Trojan insertion within legitimate library cells without modifying the synthesized netlist structure.

\subsection{Selection of Feasible Triggers}

After identifying candidate cells, our proposed framework determines whether the rare internal condition can serve as a practical Trojan trigger. Trigger conditions are validated to ensure they are functionally reachable under legal circuit operation while maintaining a very low activation probability. Candidate cells whose rare conditions are unreachable or redundant are discarded. The remaining validated cells constitute a pool of feasible Trojan insertion locations that preserve the intended functionality during normal execution while providing highly stealthy activation conditions.

\subsection{Construction of Cell-embedded Trojans}

For each validated cell, the automated framework embeds a hardware Trojan directly within the implementation of a standard-cell by modifying its internal logic rather than inserting additional gates into the synthesized netlist. The proposed Trojan is implemented as a configurable sequential sequence detector embedded inside the selected library cell. The sequence detector monitors the cell's internal rare node activations and triggers the malicious payload only after observing a user-specified sequence of events as shown in Fig. \ref{fig:cell}. By allowing arbitrary trigger sequences, the framework can generate diverse sequential Trojans with varying activation complexity and stealth characteristics. Since all modifications remain confined to the standard-cell implementation, the synthesized netlist continues to instantiate the original library cell, making the Trojan difficult to detect using conventional gate-level structural or functional detection techniques. Furthermore, the proposed framework supports the insertion of multiple independent Trojans (N) throughout the design according to user-defined configuration requirements.

\subsection{Trojan Validation}

The final stage validates every generated Trojan to ensure both stealth and correctness. The framework first verifies that the trigger condition is reachable and that the payload can propagate to at least one primary output or architecturally observable point. Functional equivalence is then evaluated under normal operating conditions to confirm that the compromised cell behaves identically to the original library cell unless the intended trigger sequence occurs. Only Trojans satisfying trigger reachability, payload observability, and functional correctness are retained as valid cell-resident hardware Trojan benchmarks.

\section{Experimental Results}

The implementation of our proposed HT insertion framework is compatible with the conventional EDA flows. We extensively evaluate our proposed Trojan insertion framework using four benchmarks with different sizes and functionalities, as shown in Table \ref{tab:overhead}. We use the GSCL 45nm PDK as the candidate standard-cell library for embedding Trojans and validating them. We use Synopsys TestMax ATPG to verify the functional validity of the trigger conditions.

\begin{table}[!ht]
\centering
\caption{Area and power overhead incurred by the Trojan inserted designs generated through our proposed framework with varying numbers of infected library cells}
\label{tab:overhead}
\resizebox{0.9\columnwidth}{!}{%
\begin{tabular}{|c|c|c|cc|}
\hline
\multirow{2}{*}{\textbf{Design}} & \multirow{2}{*}{\#\textbf{gates}} & \multirow{2}{*}{\#\textbf{infected\_cells}} & \multicolumn{2}{c|}{\textbf{Overhead (in\%)}}         \\ \cline{4-5} 
                        &                          &                                   & \multicolumn{1}{c|}{\%\textit{area}} & \%\textit{power} \\ \hline
\multirow{2}{*}{UART}   & \multirow{2}{*}{1990}    & 10                                & \multicolumn{1}{c|}{1.82}   & 1.49    \\ \cline{3-5} 
                        &                          & 20                                & \multicolumn{1}{c|}{3.63}   & 2.97    \\ \hline
\multirow{2}{*}{S38584} & \multirow{2}{*}{9785}    & 10                                & \multicolumn{1}{c|}{0.87}   & 0.47    \\ \cline{3-5} 
                        &                          & 20                                & \multicolumn{1}{c|}{1.51}   & 0.94    \\ \hline
\multirow{2}{*}{SHA}    & \multirow{2}{*}{7309}    & 10                                & \multicolumn{1}{c|}{1.15}   & 0.91    \\ \cline{3-5} 
                        &                          & 20                                & \multicolumn{1}{c|}{2.31}   & 1.81    \\ \hline
\multirow{2}{*}{AES}    & \multirow{2}{*}{160756}  & 10                                & \multicolumn{1}{c|}{0.03}   & 0.02    \\ \cline{3-5} 
                        &                          & 20                                & \multicolumn{1}{c|}{0.05}   & 0.04    \\ \hline
\end{tabular}%
}
\end{table}

\subsection{Experimental Results}

As summarized in Table~\ref{tab:overhead}, the proposed library cell Trojan insertion framework introduces minimal area and power overhead across all benchmark designs. Although increasing the number of infected cells from 10 to 20 approximately doubles the overhead within the same design, the overall impact remains very small. More importantly, the relative overhead decreases significantly as the design size increases. For example, the UART benchmark (1990 gates) incurs up to 3.63\% area and 2.97\% power overhead when 20 cells are infected, whereas the AES design (160756 gates) exhibits only 0.05\% area and 0.04\% power overhead under the same configuration. Similarly, the SHA benchmark shows only 2.31\% area and 1.81\% power overhead despite infecting 20 cells. This trend demonstrates that the cost of embedding Trojans inside standard-cell implementations becomes increasingly negligible for larger designs, indicating that the proposed automated HT generation framework scales well to realistic industrial circuits while maintaining a highly stealthy hardware footprint.

\begin{figure}[!htbp]
\centering
\includegraphics[width=\columnwidth]{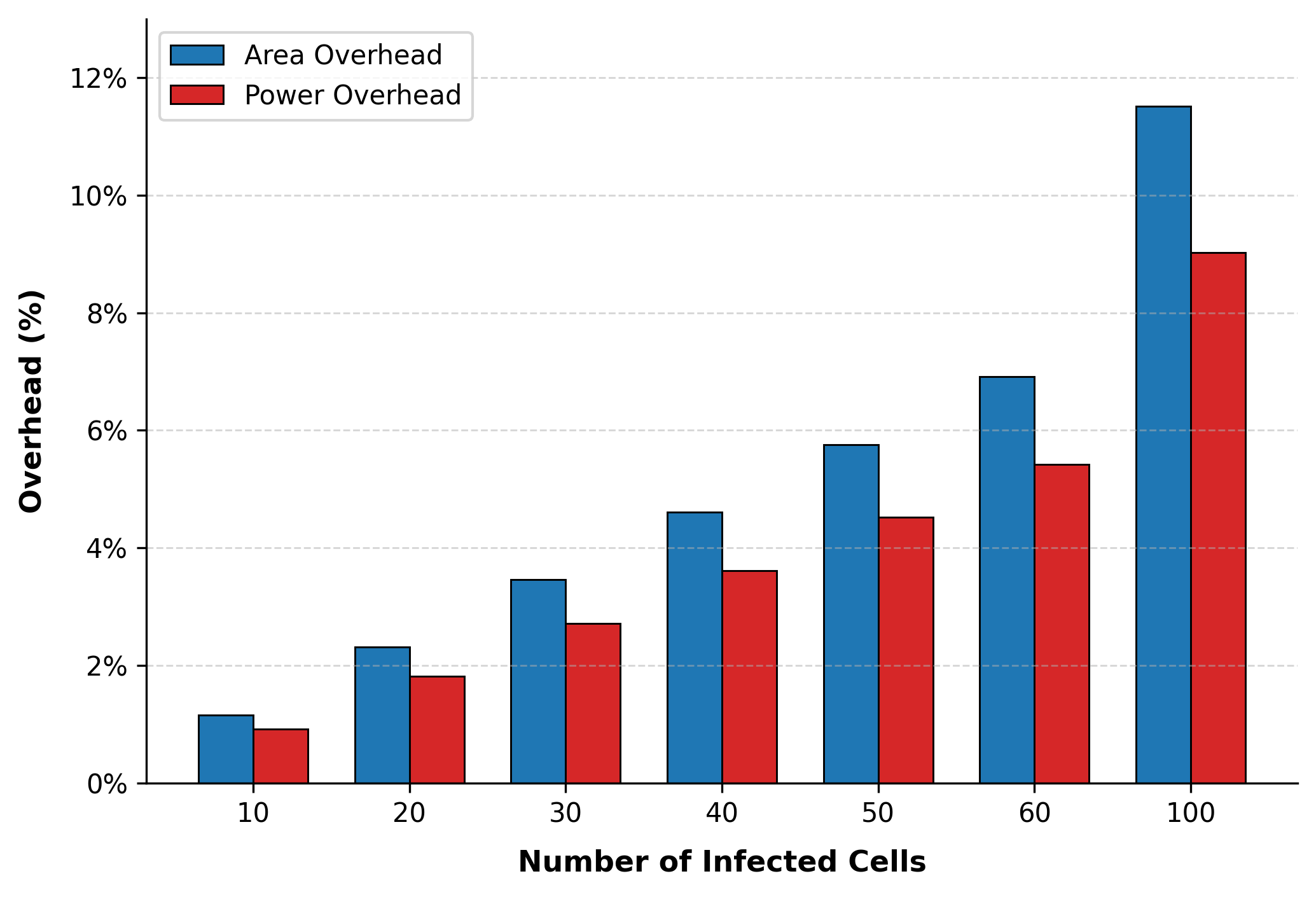}
\caption{Trend of area and power overhead for SHA design as the number of Trojan-infected cell instances is varied between 10 and 100.}%
\label{fig:trend}%
\end{figure}

Fig.~\ref{fig:trend} illustrates how the proposed HT insertion framework scales with the number of infected cells for the SHA design. As expected, both area and power overhead increase approximately linearly with the number of infected cells since additional sequential Trojan circuitry is embedded within more standard-cell instances. Even with 100 infected cells, the overhead remains modest at approximately 11.5\% area and 9\% power, while infecting only 10–20 cells incurs less than 2.5\% area and 2\% power overhead. This demonstrates that our proposed framework enables users to trade off Trojan coverage against implementation overhead by controlling the number of compromised cells. In practical attack scenarios, where only a small fraction of library cells are infected, the resulting hardware footprint is minimal, thereby enhancing the stealthiness of the inserted Trojans.

\subsection{Stealthiness Evaluation}

To evaluate the stealthiness of the generated cell-embedded Trojans through our proposed framework, we assess whether representative state-of-the-art hardware Trojan detection techniques can successfully identify the inserted modifications, as shown in Table \ref{tab:stealthy}. First, we perform extensive random simulations using a large set of randomly generated test vectors (up to 100K) to determine whether the Trojan trigger is activated during conventional functional testing. We then evaluate ATPG-based statistical activation methods, including MERO \cite{mero} and LATENT \cite{latent}, which explicitly target rare trigger conditions. We also evaluate ML-based HT detection frameworks such as \cite{gnn4tj,vipr}, which classify hardware Trojans using structural graph features extracted from gate-level netlists. Additionally, we employ Logic Equivalence Checking (LEC) \cite{clip} using commercial EDA tool \cite{conformal} to verify the functional equivalence between the HT-inserted and HT-free netlists. Unlike conventional RTL or gate-level Trojan insertion, our proposed framework embeds the malicious behavior inside the implementation of standard cells while leaving the synthesized netlist unchanged. Consequently, the infected benchmarks exhibit the same gate-level structure as the clean design, preventing structural learning methods from distinguishing compromised cells. Similarly, statistical test-generation techniques activate rare nets in the synthesized design but cannot expose behavioral changes hidden within the internal implementation of library cells. These results demonstrate that existing detection techniques are fundamentally designed for explicit gate-level Trojans and are unable to detect this new class of cell-embedded hardware Trojans.

\begin{table}[!ht]
\centering
\caption{Stealthiness evaluation of the generated HTs}
\label{tab:stealthy}
\resizebox{\columnwidth}{!}{%
\begin{tabular}{|c|cccccc|}
\hline
\multirow{3}{*}{Design} & \multicolumn{6}{c|}{HT Detection Techniques}                                                                                                                                                                                                                                                                                             \\ \cline{2-7} 
            & \multicolumn{3}{c|}{Random   Simulation}                                        & \multicolumn{1}{c|}{\multirow{2}{*}{\begin{tabular}[c]{@{}c@{}}Statistical \\ Analysis \cite{mero,latent}\end{tabular}}} & \multicolumn{1}{c|}{\multirow{2}{*}{\begin{tabular}[c]{@{}c@{}}ML-based \\ Methods \cite{gnn4tj,vipr}\end{tabular}}} & \multirow{2}{*}{\begin{tabular}[c]{@{}c@{}}LEC \\ Analysis \cite{conformal}\end{tabular}} \\ \cline{2-4}
            & \multicolumn{1}{c|}{10K} & \multicolumn{1}{c|}{50K} & \multicolumn{1}{c|}{100K} & \multicolumn{1}{c|}{}                                                                                 & \multicolumn{1}{c|}{}                                                                                &                                                                          \\ \hline
UART                    & \multicolumn{1}{c|}{\xmark}    & \multicolumn{1}{c|}{\xmark}    & \multicolumn{1}{c|}{\xmark}     & \multicolumn{1}{c|}{\xmark}                                                                                 & \multicolumn{1}{c|}{\xmark}                                                                                &  \xmark                                                                        \\ \hline
S38584                  & \multicolumn{1}{c|}{\xmark}    & \multicolumn{1}{c|}{\xmark}    & \multicolumn{1}{c|}{\xmark}     & \multicolumn{1}{c|}{\xmark}                                                                                 & \multicolumn{1}{c|}{\xmark}                                                                                &  \xmark      \\ \hline
SHA                    & \multicolumn{1}{c|}{\xmark}    & \multicolumn{1}{c|}{\xmark}    & \multicolumn{1}{c|}{\xmark}     & \multicolumn{1}{c|}{\xmark}                                                                                 & \multicolumn{1}{c|}{\xmark}                                                                                &  \xmark      \\ \hline
AES                    & \multicolumn{1}{c|}{\xmark}    & \multicolumn{1}{c|}{\xmark}    & \multicolumn{1}{c|}{\xmark}     & \multicolumn{1}{c|}{\xmark}                                                                                 & \multicolumn{1}{c|}{\xmark}                                                                                &  \xmark      \\ \hline
\end{tabular}%
}
\end{table}

\section{Conclusion}
We presented an automated framework for generating stealthy hardware Trojans by embedding configurable sequential Trojans within compromised standard-cell implementations. Unlike conventional Trojan generation techniques, the proposed approach enables attackers to covertly implant malicious functionality inside library cells while preserving the synthesized gate-level netlist, allowing the Trojans to remain dormant under normal operation and evade existing detection assumptions. We demonstrated the applicability of the framework across benchmark designs of varying sizes while incurring minimal area and power overhead. Furthermore, we evaluated the generated Trojans against representative state-of-the-art hardware Trojan detection techniques and found that none of them successfully detected the inserted cell-resident Trojans. These results expose a previously overlooked blind spot in current hardware Trojan detection methodologies and highlight the need for future security mechanisms capable of validating compromised standard-cell libraries.

\bibliographystyle{IEEEtran}
\bibliography{IEEEabrv,references}

\end{document}